\documentclass[runningheads]{llncs}
\usepackage[utf8]{inputenc} % allow utf-8 input
\usepackage[T1]{fontenc}    % use 8-bit T1 fonts
\usepackage{hyperref}       % hyperlinks
\usepackage{url}            % simple URL typesetting
\usepackage{booktabs}       % professional-quality tables
\usepackage{amsfonts}       % blackboard math symbols
\usepackage{nicefrac}       % compact symbols for 1/2, etc.
\usepackage{microtype}      % microtypography

\hypersetup{
    colorlinks,
    linkcolor={blue!80!black},
    citecolor={blue!80!black},
    urlcolor={blue!80!black}
}

\usepackage{comment}
\usepackage{graphicx}
\usepackage[table,xcdraw]{xcolor}
\usepackage{tcolorbox}
\usepackage{import}
\usepackage{macros}
\begin{document}
\title{\pname\thanks{Position paper: To appear in the 2023 Accelerating Space Commerce, Exploration, and New Discovery (ASCEND) conference, Las Vegas, Nevada, USA.}}
\titlerunning{\pnameshort}
\author
    {Tim Ellis\inst{1}\orcidID{\href{https://orcid.org/0009-0002-2315-2325}{0009-0002-2315-2325}} \and
    Briland Hitaj\inst{1}\orcidID{\href{https://orcid.org/0000-0001-5925-3027}{0000-0001-5925-3027}} \and \\
    Ulf Lindqvist\inst{1}\orcidID{\href{https://orcid.org/0009-0002-5941-0947}{0009-0002-5941-0947}} \and
    Deborah Shands\inst{1}\orcidID{\href{https://orcid.org/0000-0002-3408-8106}{0000-0002-3408-8106}} \and
    Laura Tinnel\inst{1}\orcidID{\href{https://orcid.org/0000-0002-1779-1297}{0000-0002-1779-1297}} \and 
    Bruce DeBruhl\inst{1,2}\orcidID{\href{https://orcid.org/0000-0003-1148-6010}{0000-0003-1148-6010}} 
    }
% Deborah: 0000-0002-3408-8106
% Tim: 0009-0002-2315-2325
% Briland: 0000-0001-5925-3027
% Ulf: 0009-0002-5941-0947
% Laura: 0000-0002-1779-1297
% Bruce: 0000-0003-1148-6010
%
\authorrunning{T. Ellis et al.}
\institute{
    Computer Science Laboratory, SRI International \and
    Computer Science and Software Engineering Department \\ California Polytechnic State University \\
    \email{\{firstname.lastname\}@sri.com}}
\maketitle              % typeset the header of the contribution
%
% !TEX root = ../main.tex
\begin{abstract}
Space systems enable essential communications, navigation, imaging and sensing for a variety of domains, including agriculture, commerce, transportation, and emergency operations by first responders. Protecting the cybersecurity of these critical infrastructure systems is essential. While the space environment brings unique constraints to managing cybersecurity risks, lessons learned about risks and effective defenses in other critical infrastructure domains can help us to design effective defenses for space systems. In particular, discoveries regarding cybersecurity for industrial control systems (ICS) for energy, manufacturing, transportation, and the consumer and industrial Internet of Things (IoT) offer insights into cybersecurity for the space domain. 
This paper provides an overview of ICS and space system commonalities, lessons learned about cybersecurity for ICS that can be applied to space systems, and recommendations for future research and development to secure increasingly critical space systems.

\keywords{Critical Infrastructure Security \and Industrial Control Systems \and Operational Technology \and Cybersecurity \and Space Applications.}

\end{abstract}
% !TEX root = ../main.tex
\section{Introduction}
\label{sec:introduction}

We increasingly rely on space systems for functions such as communications, navigation, and imaging to support everyday life-critical ground applications.  Space systems have become part of our critical infrastructure in domains such as energy, transportation, agriculture, and emergency services. Malicious cyber attacks against critical infrastructure systems have increased dramatically in recent years and space systems are becoming attractive targets for attackers. While space systems have unique characteristics associated with the space environment, they also share many characteristics with industrial control systems (ICS) that manage much of our ground-based critical infrastructure. Lessons learned from efforts to secure ICS may help us to understand cybersecurity risks and design effective protections for space systems. 

Space systems and ground-based critical infrastructure systems for gas, oil, and electrical power share many challenging requirements that have resulted in common system designs.  Requirements such as harsh deployment environments, remote operations, high reliability, and long-lived deployments have resulted in similar high-level system architectures and reliance on special-purpose hardware and software that supports unique analog functionality and often suffers from significant resource limitations. Cybersecurity risks that affect ICS often have analogous risks for space systems, and the lessons learned in developing security protections for ground-based ICS can help us design effective protections for space systems.  

This paper is organized as follows: Section~\ref{sec:commonalities} describes common characteristics between ICS and space systems that impact security protections and constrain how protections can be implemented or deployed. Section~\ref{sec:lessons-learned} discusses lessons learned from securing ground-based ICS, including challenges and approaches. Section~\ref{sec:new-approaches} considers how new security techniques, including some based on machine learning, may be applied to space systems and provides thoughts for further research toward securing space systems. Section~\ref{sec:conclusion} concludes with a summary of the paper.

\section{ICS and Space System Commonalities} \label{sec:commonalities}

\begin{figure}[ht]
  \centering
  \includegraphics[width=\textwidth]{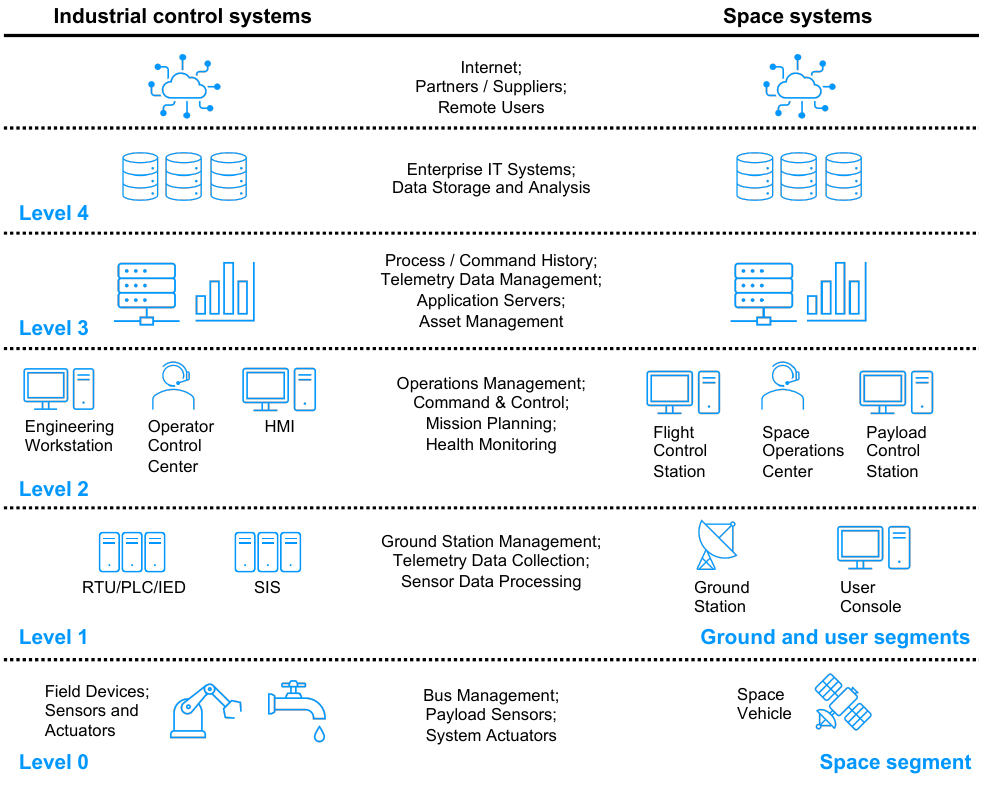}
  \caption{Space systems and industrial control systems have analogous control layers and therefore similar cybersecurity threat and mitigation opportunities.}
  \label{fig:systems}
\end{figure}

ICS for critical infrastructure and space systems share many characteristics that impact security protections, constraining how these protections can be implemented or deployed. As shown in Figure \ref{fig:systems}, the ICS control layers are similar to the space system segments, where ``mission control'', or ``operation centers'', control and monitor the remote cyber-physical systems.

{\bf System Architecture:} In large ICS, human system controllers interact with command and control (C2) systems or centers that are often designed and implemented independently from the devices operating in the field. Similarly, a ground-based mission control or operations center hosts C2 systems that remotely manage spacecraft. C2 systems are often well-resourced with much of the same type of equipment (e.g., rack-mounted servers, network-attached storage systems) found in a modern enterprise computing environment.  

Both ICS and space mission C2 systems issue commands to initiate operations on remote devices. In both the ICS and space domains, the complexity of the operations that may be initiated is increasing and newer remote devices can operate with more autonomy than older devices. Remote devices communicate telemetry back to the C2 system, which provides a view of the health of the remote system to both human operators and automated monitoring systems.

The system architecture of an ICS offers both opportunities and constraints for the design of system security protections. Remote devices are often inaccessible to operators, so capabilities for monitoring the health and behavior of devices are designed into the remote devices and communication protocols.   Security monitoring (e.g., periodic remote attestation of firmware integrity) must fit within the established remote monitoring model as well as the limited system resources. Fortunately, C2 systems typically have sufficient system resources or can be expanded to support compute-intensive analysis, such as intrusion detection.  Security protection approaches that can be designed to operate in such asymmetrical system architectures are likely to be usable in both ICS and space systems.
	
 {\bf Significant Resource Limitations and Long-lived Equipment:} Space vehicles and remote industrial equipment are often severely constrained by their computing, memory, and communications resources. Power limitations are common in such systems, where solar power is limited or batteries must be conserved.   Such physical limitations restrict the computational and communications capabilities of the systems, as processors and radios require power proportional to their performance. Memory restrictions in these systems may result from hard limits on the size of the device: mass limitations are typical for space vehicles, while industrial systems may have other dimensional limitations.

 Because both space vehicles and remote industrial equipment are difficult and expensive to deploy into remote locations, these systems are expected to operate continuously without human touch for extended periods of time: a decade or more of expected operation is not uncommon. Hardware deployed 10+ years ago with state-of-the-art processors and considerable memory capacity is operating by today's standards with obsolete processors and very limited memory. 

Resource limitations and outdated remote equipment have important impacts on system security. Updating remote device software to enable improvements in security monitoring is often impossible, as the equipment lacks the capacity to both perform its intended function and to run ancillary security software. When a security vulnerability is detected, software must be updated or patched. For remote industrial equipment, that requires either a human touch (e.g., loading new software via a thumb drive) or remote transmission of the software patch. Newer ICS and on-board space systems are designed to enable remote patching, however, slow network transmission rates can limit the frequency and size of patches. And, in both cases, there is always the risk of causing an irreversible loss of functionality if anything goes wrong with the update or its transmission.
 
{\bf Essential Analog Functionality:} Both ICS and space systems implement essential analog functions (e.g., providing sensor data, controlling valves, moving solar panels, controlling communications antennas). Security mechanisms deployed onto such systems must, themselves, do no harm by impeding these essential functions. For example, a device that must meet hard real-time requirements will not be able to execute security software that runs intermittently, consuming an unpredictable number of processor cycles. 

Essential analog functionality is both a target for attackers and a monitoring opportunity for defenders. Attackers may attempt to impede the device from performing an essential function by corrupting software or data on either the remote device or within the C2 system. Defenders can monitor data from analog functions, separate from the often more vulnerable digital channels, to assess the health and security of the remote device.

{\bf Special-purpose Software and Hardware:} Space systems and remote industrial machinery are often built from scratch with special-purpose hardware and software, which is typically necessary either to address the essential analog functionality (e.g., real-time requirements) or to withstand the harsh environments in which they must operate (e.g., space vehicle exposure to cosmic radiation, factory machinery exposure to intense vibration). Defenders are at a disadvantage when protecting systems with special-purpose components because industry best-practice techniques for enterprise security (e.g., default ``secure'' configurations for common system software) are rarely applicable and third-party security solutions do not exist. 

\begin{figure}[ht]
  \centering
\includegraphics[width=0.85\textwidth]{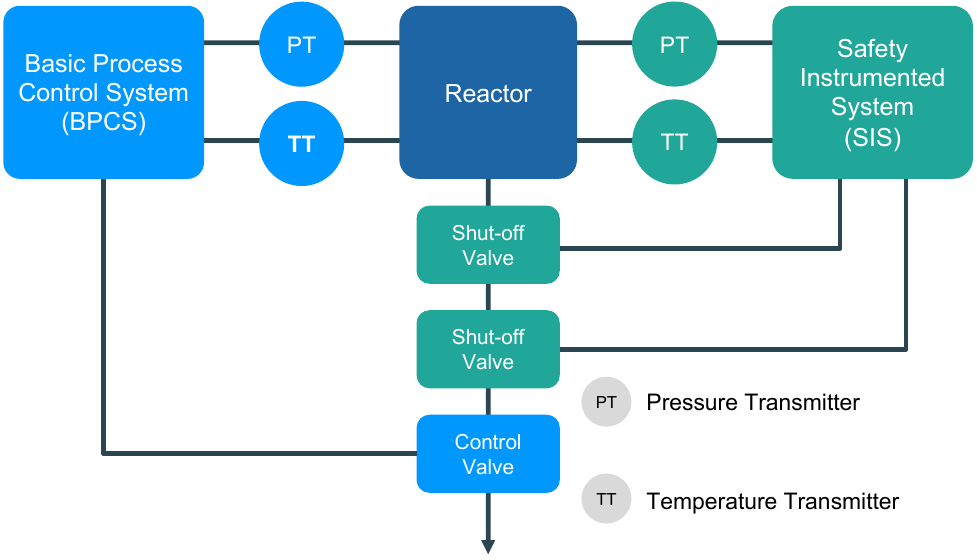}
  \caption{Example safety instrumented system installed in parallel with process control system to bring the system to a safe state when predetermined thresholds are violated~\cite{NIST-SP-800-82r3}.}
  \label{fig:SIS_figure}
\end{figure}

{\bf Safety Instrumented Systems:} The ICS industry developed and adopted many fundamental building blocks and principles for ICS functional safety from 1980 to the mid-1990s. IEC standard 61508 called for using risk-based functional safety integrity levels (SILs) to protect ICS. Initially, process control systems (PCS) implemented these safety functions in addition to automating normal operations. In the early 2000s, the industry introduced IEC standard 61511, which called for separating ICS safety control functions from the PCS and using safety instrumented systems (SIS). SIS are specialized PCS that use dedicated sensors and control logic to monitor the operations of industrial plants for dangerous process conditions (e.g., pressure out of safe range) that could result in hazardous and potentially catastrophic incidents (Figure \ref{fig:SIS_figure}). When unsafe conditions arise, the SIS automatically adjust process components (e.g., valve position) to maintain safe operations. Correct SIS operation wholly depends on the integrity and availability of its sensors, control logic, and actuators. Over time, these ICS environments have increasingly incorporated commodity information technology (IT) systems to provide operator workstations, historian databases, and other IT-like functions. Asset owners have also adopted IT networking approaches and equipment, including specialized TCP/IP-based protocols, to enable IT-based systems to communicate with operational technology (OT) systems. These IT systems and networks bring all their inherent vulnerabilities into the OT environments and present a broad attack surface and attack paths into the OT environments. Furthermore, legacy OT systems and devices were not designed with human attackers in mind.

Spacecraft rely similarly on their sensors, control logic, and actuators to maintain the safety and correct vehicle and payload operations. They also rely on ground operations, composed of IT systems and networks, to provide spacecraft command and control and receive system updates. As with ICS and SIS, attackers can leverage any IT systems and networks in use to affect spacecraft sensors, control logic, and actuators. The actual impact of such attacks depends on each component's usage. For example, spoofed sensors, corrupted control logic, or unavailable actuators could result in a spacecraft or payload's failure to perform its intended function or the spacecraft prematurely descending into the earth's atmosphere and burning up. Such issues could adversely affect other operations that depend on the services provided by the spacecraft.

{\bf Extreme Availability Requirements:} Many ICS and space systems are mission- and/or safety-critical, e.g., electrical power and natural gas distribution systems, and space-based navigation systems. Security protections are essential to support high availability. Unfortunately, the availability requirements limit or exclude the use of some common security mitigations. For example, operators are rarely able to take system services offline to install a software patch to fix a vulnerability. Security protections for both ICS and space systems must be designed to address high availability requirements for critical systems.

While ICS and space systems have some significant implementation differences due to their distinct deployment environments, their many commonalities, as summarized in Table \ref{ICS-Space-Similarities}, encourage us to examine how security protections designed for one type of system may apply or at least provide non-obvious security enhancements to the other. In the next section, we consider how lessons learned from efforts to secure ICS may apply to space systems. 

\begin{table}[ht]
\centering
\caption{ICS has many similarities with space systems (highlighted) related to resource contraints, and a few differences related mostly to physical access.}
\label{ICS-Space-Similarities}
\begin{tabular}{|l|l|l|}
\hline
\multicolumn{1}{|c|}{\textbf{System Characteristics}} & \multicolumn{1}{c|}{\textbf{Space Systems}} & \multicolumn{1}{c|}{\textbf{ICS and SIS}}  \\ \hline
Control console                                       & \cellcolor[HTML]{DAE8FC}IT, well resourced  & \cellcolor[HTML]{DAE8FC}IT, well resourced \\ \hline
Device physical access                                & Mostly none                                 & Limited                                    \\ \hline
Device computing                                      & \cellcolor[HTML]{DAE8FC}Limited             & \cellcolor[HTML]{DAE8FC}Limited            \\ \hline
Device power                                          & Limited (battery, solar)                    & Mostly grid + battery/UPS                  \\ \hline
Link bandwidth                                        & \cellcolor[HTML]{DAE8FC}Varies              & \cellcolor[HTML]{DAE8FC}Varies             \\ \hline
Availability requirements                             & \cellcolor[HTML]{DAE8FC}High                & \cellcolor[HTML]{DAE8FC}High               \\ \hline
Device replacement cost                               & High                                        & Varies                                     \\ \hline
Durability                                            & \cellcolor[HTML]{DAE8FC}High                & \cellcolor[HTML]{DAE8FC}High               \\ \hline
\end{tabular}%
\end{table}

% !TEX root = ../main.tex
\section{Lessons Learned}
\label{sec:lessons-learned}

In security over the past 50 years or so, we have learned the hard way that to have a chance of not failing we must consider a whole system in context, including its components, interfaces, emergent properties, applications, environment, people, and processes~\cite{Anderson1994WCF,neumann2006holistic}. 
Effective security protections must holistically address a system's components, interfaces, emergent properties, applications, and administrative processes in addition to the system environment and human operators. %

Legacy ICS for critical infrastructure were not often well-protected against cybersecurity attacks. Machinery deployed into remote locations was not accessible via the Internet, so ICS were designed against minimal cybersecurity requirements. As cost and efficiency considerations have driven increasing network connectivity for both C2 and remote ICS machinery, these systems now face much more challenging requirements for secure operations. ICS cybersecurity risks and protection methods are now actively studied by researchers, and protection tools and techniques are being deployed into modern ICS.  In this section, we discuss lessons learned from both research and practice in cybersecurity for ground-based ICS and consider their applicability to space systems.

\subsection{Challenges of Applying IT Security to OT Systems}

ICS environments include a mix of special-purpose OT components and relatively standard IT components. OT components often include specialized devices that implement essential analog functionality and PCS that control these devices. IT components include operator consoles (which typically run Microsoft Windows), servers that run standard IT services (e.g., directory services, database management systems), and common enterprise networking equipment (e.g., switches, routers). Implementing modern IT system security controls is essential to protect the IT components in these mixed IT-OT environments, as IT systems and insecure communications can be the weakest links for cyber attacks against OT systems~\cite{Tinnel2022}. As discussed in Section~\ref{sec:commonalities}, defenses developed for IT-only systems that depend on frequent software patches, standard secure configurations, and third-party security products are often not applicable, or even viable, for the OT components of mixed IT-OT ICS. OT-specific security products and measures must be developed and integrated to effectively reduce cybersecurity risks.

One example of OT-specific security approaches that may be adapted for space is the use of one-way data gateways to isolate the OT system from the IT infrastructure. In these system architectures, a digital emulator of the OT system is used on the IT side of the OT-IT boundary. Sensor data is allowed to flow from the OT side to the IT side to allow the emulators to provide real-time visibility into the actual OT system operations. But, under normal operating modes, control data is not allowed to flow from the IT side to the OT side of the boundary. Thus, even if the IT side is compromised, access and malicious control of the OT system is blocked. This control flow can still be allowed in specific and limited maintenance mode operations to permit control adjustments when necessary while limiting vulnerabilities during normal operations.

\lesson{Cybersecurity controls designed for IT systems are often insufficient to protect systems that include both IT and OT components. Specialized OT cybersecurity controls are also needed.}

ICS operators are application domain experts and may have IT system training but rarely have specialized cybersecurity expertise. They rely on vendors to deliver securely configured system components. Operators are frequently unaware of security options or are not trained to configure them. Operators may delay the application of even security-critical software patches, due to concerns that a patch may degrade or break system functions that could result in a costly outage. Secure ICS operations teams require embedded cybersecurity expertise and at least basic cybersecurity training for all operators and managers. For example, basic understanding of cybersecurity threat identification and tactics could help system operators better understand where and how attack surface vulnerabilities may occur. This, in turn, provides a better understanding of the awareness and criticality of using cybersecurity assessments and hygiene when installing, replacing, or configuring ICS system components to reduce or mitigate possible vulnerabilities. 

\lesson{Operations teams should be augmented to include cybersecurity expertise and training.}

Specialized devices and process control systems often lack basic security features, are difficult to configure, or cannot be configured securely. For example, serially connected devices typically have no concept of authentication and trust and will execute any command received. Additionally, these devices often include undocumented maintenance commands and toggle commands that require no knowledge of the parameters affected. These can be easily abused to malicious effect.
Many device communication protocols were not designed with security in mind and can be easily manipulated or forged~\cite{Tinnel2021HART}, and device controllers often do not authenticate the source of remotely received device configuration commands. Software patch frequency for OT device controllers is much lower than for typical IT systems and services, so vulnerabilities tend to persist for a longer time. In addition, found vulnerabilities are often held close to the vest and patches are not issued at the same frequency as in the IT domain.

One example of cybersecurity design techniques that could be adapted for space is the application of language-theoretic security, or LangSec, analysis. LangSec analysis treats the data that a system operates on as part of the computing machinery, and therefore should be treated as a formal language that must be parsed and validated before operating on~\cite{Sassaman2013LangSec}. From this perspective, system designers need to consider the intra-component communication channels and ensure that the input data at each application programming interface (API) is validated prior to use. Further, these communication APIs should avoid context specific protocols where assumptions are made based on one set of inputs about how to handle a subsequent set of inputs. This can lead to malicious attackers setting up false contexts that can then be exploited with specially crafted input data streams. This constraint can impose possible performance challenges, where lengthy data protocols often allow the recipient to begin processing the input data prior to fully receiving all the input to minimize response latencies. However, this approach can lead to the vulnerability noted above. Shorter, context-free communication protocols are much easier to validate and secure the system.

\lesson{Additional external cybersecurity protections are necessary for legacy systems with OT components. New systems should be designed and analyzed to ensure they meet modern cybersecurity requirements.}

\subsection{Software Supply Chain Security}

Both ICS and space systems have a complex mix of hardware, firmware, software, and operating parameters that must successfully integrate to achieve the desired mission. 
From the Thompson compiler~\cite{thompson1984reflections} to the Solarwinds hack~\cite{peisert2021perspectives} it has been obvious that it matters what software an organization uses to develop its own system. It is also evident that having a clear and usable view of the products one is using as a foundation is critical given the velocity of new vulnerability discovery and patch releases.

Furthermore, modern software applications integrate code from a diversity of libraries and tools including compilers, plugins, and integrated development environments (IDEs). It has become increasingly clear that this software must be understood and managed in order to mitigate risk in a system. This has lead to recent pushes by the U.S.\@ government for the use of software bill of materials (SBOM) for all software sold to the  government \cite{wh2021exec-order,dhs2023newsrelease}. Generally, SBOMs will include robust information about the vendor, version, dependencies, and relationship between any software used in a project. If all suppliers have a similar SBOM available, a software project can generate a complete dependency and version graph to have a clear understanding of the exact software that is incorporated into their environment.  As vulnerabilities and patches are discovered this can allow for rapid risk assessment and patching as appropriate. For ICS, this also includes device-specific software and firmware such as programs for field-programmable gate arrays (FPGAs) and logic programs for programmable logic controllers (PLCs).
It is also essential for developers to have continuous integration and testing in place to verify the use of libraries and functions, mitigating the risk that the update will break the code base. 

ICS often require extreme specialization of components which historically would have been developed using high-cost application-specific integrated circuits (ASICs). However, we have recently seen an increasing trend and adoption of FPGAs for lower-cost, highly adaptable, and fast processing.  For example, FPGAs have been adopted into logic controllers for industrial control systems and radios in software-defined radios. While FPGAs have tremendous utility, their software often can be challenging to understand, parse, and test. Therefore, it is critical that low-level FPGA code be included in the SBOM and analysis of secure software.

\lesson{All software dependencies must be known and traceable. Software integrity awareness and protection is needed throughout the lifecycle.}

\subsection{Hardware Supply Chain Security}

The possibility of hardware tampering and Trojans has also been acknowledged in the global supply chain. However, the possible existence of hardware Trojans could lead to security failures in availability, confidentiality, and integrity. Numerous scenarios could be imagined for ICS, including ransomware, backdoors, acts of war, and beyond that could easily motivate an attacker to design a hardware Trojan.

Therefore, similar to the need for SBOM, it is essential that a complete listing of expected hardware designs, components, versions, and changes be available for verification purposes. Having a set of printed circuit board (PCB) designs allows for automated visual inspection of boards and components to provide a base level of assurance. Furthermore, using standard testing techniques, components can be tested before assembly and integration to verify that they are functioning as expected. 

Once assembly is completed, hardware modules often move through many locations for testing, imaging, integration, and deployment. ICS modules could be intercepted, tampered with, and forwarded---making a previously verified system no longer secure.  Therefore, for any system that is going to be out of the operators' hands, it is beneficial to explore anti-tamper, or at least tamper-detection, solutions. 

\lesson{All hardware dependencies must be known and traceable. Hardware integrity protection is needed throughout the lifecycle.}

\subsection{Software Update Challenges}

Updates, including over-the-air (OTA) updates of software, firmware, FPGA images, and configuration files can help mitigate security vulnerabilities by continually patching and removing code that is found to be insecure. However, the same mechanisms can be a potential vulnerability if not appropriately secured.

For OTA updates in automobiles, the National Highway Traffic Safety Administration (NHTSA) recommends that the integrity of software updates should be verified from the source, to the delivery, and to the install~\cite{national2020cybersecurity}. This is critical in any hard-to-reach or critical system because a bad update could have high cost consequences.  Software signing, appropriate key management, efficient use of cryptographic primitives, and good software engineering practices are all required to securely distribute software. Weaknesses or negligence in any of these domains can lead to critical failures and loss of functionality of a high-value asset in ICS. 

\lesson{Updates need strict integrity and access controls from source to installation.}

\subsection{Emergent Physical Properties}

A cyber-physical system (CPS) often includes complex computing environments interacting with diverse sensors and actuators. Sensors and actuators can fail, drift, or be attacked. When multiple controllers or agents are integrated in a complex system, the interactions of these independent systems can have emergent and impactful properties.  In particular, a control system may be designed with assumptions about the other controllers operating nominally or with only stochastic errors. However, many control systems are not designed with adversarial interactions in mind, in which adversaries are purposely choosing inputs and parameters to cause the system to reach an unstable state with unpredictable results and possibly catastrophic failures. An early example was the Aurora Generator Test at Idaho National Laboratory in 2007, in which a 27-ton diesel-powered generator was physically destroyed by a cyberattack against a protective relay that repeatedly disconnected and reconnected the generator out of sync with the power grid~\cite{Greenberg2020Aurora}. The Stuxnet malware allegedly targeted uranium enrichment centrifuges in an Iranian nuclear facility in Natanz, first by causing overpressure and later by manipulating rotor speed~\cite{Langner2013}.  Physical sensors can also be manipulated by analog attacks, as demonstrated by Fu and Xu~\cite{Fu2018sensors}. Similarly in space vehicles, sensors and actuators for navigation and attitude controls could be targets for manipulation attacks to disrupt, disable, or damage the vehicle.

\lesson{Physical properties have security impacts and must be considered.}

\subsection{Safety Instrumented Systems Best Practices}

Critical systems traditionally use redundant independent components to achieve a degree of system dependability~\cite{al2009comparative}. Redundancy here is designed to tolerate random faults, such as hardware failures or radiation-induced bit flips, and is not necessarily effective against intentional attacks. Asset owners often use the same products for redundancy to reduce system design and maintenance costs. Unfortunately, homogeneity means that primary and backup components have the same vulnerabilities, allowing attackers to compromise all reachable components easily. A single supply chain attack against a component patch can compromise all like components, even with physical and management separation. 

Cybersecurity evaluations of SIS, their instruments, and their management software~\cite{logiicp12} illuminated this issue. Safety system instruments were found to have weak protections that were, in many cases bypassable, and without further protections, could be changed in authorized and malicious ways. The SIS engineering workstation (EWS) has a persistent trust relationship with the SIS and is used to configure and control the SIS and all its sensors and actuators. By compromising the weak distribution channel for instrument vendor plugins, evaluators could install malware directly on the EWS and, coupled with weak instrument protections, use its trust relationship to silently reconfigure instrument set points in ways that could cause catastrophic results. This illustrates the peril of something commonly found in OT environments---persistent trust relationships between components. The emerging notion of zero-trust architectures aims to reduce such persistent trust relationships by applying the traditional security principles of least privilege and complete mediation~\cite{Saltzer:1975:PICS}. 

These cybersecurity evaluations also yielded other lessons learned and recommendations, many of which apply to the space domain. Below we present some key lessons from ICS cybersecurity challenges and SIS that have not already been discussed and that may improve security for space systems.
\begin{itemize}
    \item Isolate and separately protect safety and C2 portions of the system. 
    \item Use security best practices and carefully monitor and control engineering and other IT workstations that have trust relationships with critical control systems.
    \item Authenticate the source of messages sent to all OT devices.
    \item Ensure all OT devices have non-bypassable, out-of-band (or separate channel) locking mechanisms to block unauthorized change requests that could come from rogue or compromised system components. In ICS environments, some devices have physical locks but some do not. Uncrewed spacecraft need analogous locking mechanisms that are controlled separately from the main system.
    \item Identify any OT devices that lack basic security features or cannot be configured securely and apply alternative external cybersecurity protections and monitoring. 
    \item Monitor the systems and networks for unauthorized device and system reconfiguration requests.
\end{itemize}

\lesson{Redundancy and independence should be considered differently for security and safety than for general dependability.}
% !TEX root = ../main.tex
\section{ICS Security Approaches Adaptable for Space Systems}
\label{sec:new-approaches}

Following the lessons learned discussion in Section~\ref{sec:lessons-learned}, in this section we recommend a series of emerging areas that can serve as focal points when considering (near) future developments in space systems.

\subsection{Modeling, Simulation, and Analysis} \label{sec:model}

Verifying robust, secure, and resilient operations of complex CPS requires test and simulation capabilities that closely model the real operational environment. These simulation platforms, often referred to as ``digital twins'' have been used for ground-based ICS, to provide high-fidelity modeling of the physical process being controlled, the sensors used to monitor and measure its operations, the physical actuators used to manage and control the process, and the controller logic itself. Similarly, digital twin technologies have been widely used for space system modeling and must also include the kinematic environment, such as the orbital mechanics of a typical low earth orbit (LEO) constellation, and the associated dynamic network interconnections that result. This simulation environment must be capable not only of modeling any constellation configuration of satellites, orbital inclinations of the planes and stagger of the space vehicles in adjacent planes, but also to accurately model the complex communication interconnections that form the dynamic mesh network, the resulting data flows, limitations, and resulting vulnerabilities. 

Furthermore, analysis and design require the ability to vary these configuration parameters, the cross-link characteristics (e.g., optical vs radio), and a variety of network protocols and encryption methods. For example, SRI International has developed Skyline, a constellation simulator meeting these requirements, and that has been used and refined on multiple government and commercial space design and analysis programs to design, assess, and refine robust and adaptive space network communications. In addition, Skyline supports custom applications running in either the space or ground segment, on simulated or actual hardware, to allow analyses and validation of various network and security protection schemes. We encourage expanding existing research into the development of constellation simulators such as Skyline with the primary goal of designing, assessing, and refining robust, secure, and adaptive space communication networks.

\subsection{Beyond Robust Control }

Robust control over faulty networks has previously been studied in industrial control systems \cite{garone2010lqg}. The assumptions of many of these control designs center around a stochastic process with bounded noise. For example, consider a sensor monitoring flow control communicating to a controller that will maintain a target flow. The sensor will generally demonstrate stochastic variance with a predictable bound from noise. The controller designed for such a system  responds to the variations in the process to keep the system within a desired operating range while adjusting to potential network behaviors.

However, an attacker can cause targeted disturbances in a system that are outside of the controller's expected operating range~\cite{debruhl2018optimizing}. For example, an attack could target the flow controller mentioned above to cause the system to overflow in ways that are predictable for the attacker but unknown to the system designer. To mitigate potential failures, a system designer can enumerate and model the domain of possible attack controllers and related failures. Using advanced simulators, like those proposed in Section~\ref{sec:model}, it is possible to enumerate the full attack surface for a system. Since space systems often have extremely robust and high-fidelity digital twins and system models, this approach can be applied. Using this type of attack modeling, a designer can discover corner operational cases. Identifying these corner-cases can allow for hardening control systems against targeted attacks.

Furthermore, by using high-fidelity models to create digital twins and leveraging high-fidelity physics models, we can also create model-based intrusion detection systems to identify when behavior differs from the expected model behavior. Model-based intrusion detection can detect anomalies in industrial control systems~\cite{Cheung:2007:SCADAIDS}. Since space systems have high-fidelity and reliable models of their physical behavior, we can apply similar techniques. By carefully selecting inputs for the space system with known variations, the associated variations should be observable in the physical behavior of the system. This can be used to detect unexpected behaviors regardless of whether they are malicious or not.

\subsection{Automating Testing and Standardizing BOMs and SBOM Testing}

Software and hardware supply chain management is critical in both ICS and space systems. However, full manual testing of hardware or software in systems as complex as a spacecraft is nearly impossible. Emerging software and hardware bill of materials (SBOM and HBOM) generation and management tools, automated testing, and fuzzing with physical and digital twins can allow for a continuous automated testing of software across the development pipeline including firmware and configuration parameters. We therefore propose a greater standardization of hardware and software supply chain BOMs, as well as testing for security applications. These should include both traditional IT-based hardware and software, which have been extensively addressed, as well as standards for the specialized hardware, firmware, and software used in ICS and space systems.

\subsection{Secure and Privacy-Preserving ML for Space Applications}
\label{ssec:aiml}

In the previous sections, we focused on delineating similarities between ICS and space systems, highlighting cybersecurity risks in ICS, and describing potential mitigations of similar threats in the space domain. In this section, we shift the attention toward recent trends in machine learning (ML), highlight some of the recent work utilizing ML in ICS and space systems, and point out potential new threats and attack vectors, stressing the need for security and privacy by design when incorporating state-of-the-art ML applications in the space domain.

Developments in the field of ML, and particularly in deep learning (DL), have been rapidly reshaping the landscape in fields like computer vision, natural language processing, and speech recognition~\cite{otter2020survey,grigorescu2020survey,rombach2022high,gaspari2020naked}. In the context of ICS, ML has played a crucial role in addressing problems like intrusion detection~\cite{umer2022machine,kus2022false}, bot detection~\cite{jayalaxmi2022debot}, anomaly detection~\cite{abdelaty2021daics,shuaiyi2023process,jiang2022deep}, and more~\cite{koay2023machine,liakos2019machine,al2020ensemble,li2020deep}. Similarly, ML and DL applications have been widely adopted in the space domain, providing solutions to a wide variety of problems, see~\cite{nasaAI2023,esaAI2022,swope2022benchmarking,dunkel2022benchmarking,vazquez2021machine,salim2023deep,mandrake2022space,ortiz2023onboard}. We posit that adoption of ML/DL solutions in the space domain will rapidly grow, especially as these models become more resource-efficient, and concurrently with the progress being made toward explainable results and mitigating potential biases in the training data.

A growing body of work has demonstrated that ML/DL models are susceptible to a wide array of security and privacy threats, which could seriously undermine the integrity and reliability of such models. Security attacks like adversarial samples~\cite{papernot2016limitations,carlini2017towards}, Trojans~\cite{liu2017trojaning}, and backdoors~\cite{gu2019badnets} are able to compromise state-of-the-art ML/DL models, making them prone to unexpected actions and behavior. On the other hand, privacy attacks, including but not limited to property inference~\cite{ateniese2013hacking,ganju2018property}, model inversion~\cite{fredrikson2015model}, and membership inference~\cite{shokri2017membership}, can effectively leak sensitive information or determine the presence of properties and inputs present in the original training data, often considered private intellectual property given its prominent role in training of effective ML/DL models. 

The attacks listed here target different components of the ML pipeline with defenses against them still remaining an open problem despite the significant amount of work to detect, prevent, and mitigate such threats. Furthermore, new threats coming as a result of reliance on third-party, pre-trained models downloaded from unvetted repositories, often the case when a party is limited to resource constraints or limited computational data, could open the door to new attacks that not only hinder the ML/DL models but also compromise target infrastructures~\cite{hitaj2022maleficnet}. We believe that threats such as the ones listed in this section should be taken into consideration when designing forward-looking ML/DL applications for both the ICS and the space domains, emphasizing the need for incorporating additional checks on the entirety of the development pipeline. The added complexity and constraints when operating in space environments, combined with the fast-paced evolution of this area, adds an extra layer of responsibility when developing (explainable) ML/DL applications that are capable of anticipating and adapting to changing conditions that require little-to-no human assistance, and are robust to adversarial threats.
\section{Conclusion}
\label{sec:conclusion}

Space systems share so many security-relevant characteristics with ground-based cyber-physical systems that we would be remiss not to apply important lessons learned over decades of work with the latter. In this paper, we identify such commonalities and describe them in detail. Based on the common characteristics, we identify lessons learned from securing cyber-physical systems, particularly those used in industrial control settings, and we describe how those lessons could apply to space systems. The development of space systems is undergoing rapid and significant change, and the changes are similar to those experienced in the development of ground-based industrial automation systems in past decades; therefore, new approaches are needed to meet cybersecurity requirements. We describe emerging technology approaches that we believe have the potential to be highly relevant and applicable to meet the cybersecurity challenges of the space systems of the future.
\bibliographystyle{splncs04}
\bibliography{main}
\end{document}